\documentclass{epl}

\usepackage{graphicx}
\usepackage{amssymb}
\usepackage{amsfonts}
\usepackage{amsmath}

\newcommand{\tst}{\textstyle}

\newcommand{\mbf}{\mathbf}
\newcommand{\mrm}{\mathrm}

\title{Sympathetic cooling of trapped fermions by bosons in the presence of particle losses}
\shorttitle{Sympathetic cooling of trapped fermions by bosons}
\author{Z. Idziaszek\inst{1,2} \and L. Santos \inst{3} \and M. Lewenstein \inst{4}}
\shortauthor{Z. Idziaszek \etal}
\institute{
  \inst{1} Istituto Nazionale per la Fisica della Materia, BEC-INFM Trento, I-38050 Povo (TN), Italy \\
  \inst{2} Centrum Fizyki Teoretycznej, Polska Akademia Nauk, 02-668 Warsaw, Poland \\
  \inst{3} Institut f\"ur Theoretische Physik III,
Universit\"at Stuttgart, Pfaffenwaldring 57 V, D-70550 Stuttgart, Germany\\
  \inst{4} Institut f\"ur Theoretische Physik, Universit\"at Hannover, D-30167 Hannover, Germany
}

\begin{document}

\maketitle

\begin{abstract}
We study the sympathetic cooling of a trapped Fermi gas interacting with an ideal Bose gas below the critical 
temperature of the Bose-Einstein condensation. We derive the quantum master equation, which describes the 
dynamics of the fermionic component, and postulating the thermal distribution for both gases 
we calculate analytically the rate at which fermions are cooled by the bosonic atoms. The particle losses
constitute an important source of heating of the degenerate Fermi gas. We evaluate the rate of loss-induced
heating and derive analytical results for the final temperature of fermions, which is limited in the presence
of particle losses.
\end{abstract}

Evaporative cooling has proven to be  an essential tool to 
obtain degenerate Fermi gases\cite{Jin,Salomon,Hulet,Thomas,Ketterle,Inguscio}.
In Fermi systems, the $s$-wave collisions of indistinguishable 
particles are forbidden due to the antisymmetry requirement,  and therefore the only way to cool 
fermions using collisions, is to cool them  sympathetically by bringing them in contact with 
atoms in different hyperfine states, or another species.
In the regime of quantum degeneracy, however, also the collisional processes in Fermi-Fermi mixtures 
are strongly suppressed due to Pauli blocking. As a consequence,  
the efficiency of sympathetic cooling is reduced, and the particle
losses, which provide an important source of heating in the degenerate regime \cite{Timmermans},
prevent from reaching the lower temperatures T, necessary, for instance, for achievement of the superfluid BCS state. 
This problem may be avoided by adiabatic crossing of a Feshbach resonance, since very low temperatures may be achieved 
\cite{ingenious} when moving from a molecular BEC regime \cite{mBEC} to the BCS regime. In this crossover, which has recently 
attracted a large attention, strong evidences for superfluidity have been observed \cite{mBEC1,mBEC2,mBEC3,mBEC4,mBEC5}.
Another obstacle appears when a Fermionic gas is 
sympathetically cooled using a  Bose-Einstein condensate, since as a consequence of the superfluid properties 
of the condensate, the sympathetic cooling is limited to velocities larger 
than the sound velocity \cite{Timmermans1}.

Sympathetic cooling has been recently 
considered by several theoretical groups
\cite{Timmermans1,Maciek,Geist,Crescimanno,Holland1,Papenbrock,Geist1,Wouters}.
It should be stressed, however, that the limitations due to particle losses have been addressed in detail
only recently by Carr \etal \cite{Carr}, who have studied the cooling of fermions by an ideal 
homogeneous Bose gas at $T=0$. In this paper we extend the result of Ref. \cite{Carr}, and investigate 
the influence of the particle losses on the sympathetic cooling of harmonically trapped gases at finite $T$.
We consider here a simplified model which assumes that fermions are in contact with an ideal Bose gas,
which is kept at constant $T$. This can be achieved if the bosons are continuously cooled by means
of evaporative cooling or eventually when the bosonic component is much larger than the fermionic one.

The gas of $N_f$ fermions is confined in a harmonic trap of frequency $\omega_f$. In the following 
we assume that the trapping potential is spherically symmetric, however, our results can be generalized 
to anisotropic traps. The fermions interact with a gas of $N_b$ bosonic atoms, confined in 
a harmonic trap of frequency $\omega_b$. We treat the gas of bosons as a reservoir, which assumes that its 
temperature $T_b$ and statistical properties do not change during the cooling process. 
The Hamiltonian of the system reads
\begin{equation}
\label{H}
H = \sum_\mbf{n} \varepsilon_\mbf{n} a_\mbf{n}^{\dagger} a_\mbf{n} + 
\sum_\mbf{m} \varepsilon_\mbf{m}^b b_\mbf{m}^{\dagger} b_\mbf{m} + \sum_{\mbf{n},\mbf{n}^{\prime},
\mbf{m},\mbf{m}^\prime} U_{\mbf{n},\mbf{n}^{\prime},\mbf{m},\mbf{m}^{\prime}} a_\mbf{n}^{\dagger} a_{\mbf{n}^\prime}
b_\mbf{m}^{\dagger} b_{\mbf{m}^\prime},
\end{equation}
where $\mbf{n}$ and $\mbf{m}$ label the states of the three-dimensional harmonic 
oscillator, $\varepsilon_\mbf{n}$ and $\varepsilon_\mbf{m}^b$ denote single particle energies of fermions and bosons,  
respectively, and $a_\mbf{n}$ ($b_\mbf{m}$) is the creation operator of fermions (bosons) fulfilling usual 
anticommutation (commutation) relations. In the low-temperature regime considered here, the scattering between
atoms of both species is purely of $s$-wave type and can be modeled by a delta pseudopotential with scattering length 
$a$. Thus the matrix elements of the boson-fermion interactions are given by 
\begin{equation}
\label{U}
U_{\mbf{n},\mbf{n}^{\prime},\mbf{m},\mbf{m}^{\prime}} = g \int \! \upd^3r \, \psi_\mbf{n}^\ast(\mbf{r}) 
\psi_{\mbf{n}^{\prime}}(\mbf{r}) \phi_\mbf{m}^\ast(\mbf{r}) \phi_{\mbf{m}^{\prime}}(\mbf{r}),  
\end{equation}
where $\psi_\mbf{n}(\mbf{r})$ and $\phi_\mbf{m}(\mbf{r})$ are the single-particle eigenfunctions of fermions and bosons, 
respectively, $g = 2 \pi \hbar^2 a / \mu_m$, and $\mu_m= m_b m_{\!f}/(m_b + m_{\!f})$ denotes the reduced mass ($m_b$ is
the mass of boson and $m_{\!f}$ is the mass of fermion). We apply standard
techniques of the theory of quantum-stochastic processes \cite{Gardiner,Carmichael} and obtain the quantum Master
equation (ME) for the density matrix $\rho$ of the fermions \cite{Maciek}
\begin{equation}
\label{ME}
\frac{d \rho}{dt} = - \frac{i}{\hbar} \left[ 
H_F + H_{F-F}, \rho \right] +
\sum_{\mbf{n},\mbf{n}^{\prime},\mbf{m},\mbf{m}^{\prime}} 
\Gamma_{\mbf{n},\mbf{n}^{\prime}}^{\mbf{m},\mbf{m}^{\prime}} \left( 2 a_\mbf{m}^{\dagger} a_{\mbf{m}^\prime} \rho
a_\mbf{n}^{\dagger} a_{\mbf{n}^\prime} - a_\mbf{n}^{\dagger} a_{\mbf{n}^\prime} a_\mbf{m}^{\dagger} a_{\mbf{m}^\prime}
\rho - \rho a_\mbf{n}^{\dagger} a_{\mbf{n}^\prime} a_\mbf{m}^{\dagger} a_{\mbf{m}^\prime} \right),
\end{equation}
where $H_F = \sum_\mbf{n} \varepsilon_\mbf{n} a_\mbf{n}^{\dagger} a_\mbf{n}$
and the rate coefficients are of the form:
\begin{equation}
\label{Gamma}
\Gamma_{\mbf{n},\mbf{n}^{\prime}}^{\mbf{m},\mbf{m}^{\prime}} = \frac{\pi}{\hbar^2 \omega_b}
\delta(\varepsilon_\mbf{n}+\varepsilon_\mbf{m},\varepsilon_{\mbf{n}^\prime}+\varepsilon_{\mbf{m}^\prime})
\sum_{\mbf{k},\mbf{q}} U_{\mbf{n},\mbf{n}^{\prime},\mbf{q},\mbf{k}} 
U_{\mbf{m},\mbf{m}^{\prime},\mbf{k},\mbf{q}} n_\mbf{q}^b (1 + n_\mbf{k}^b) 
\delta(\varepsilon_\mbf{m}+\varepsilon_\mbf{k}^b,\varepsilon_{\mbf{m}^\prime}+\varepsilon_\mbf{q}^b).
\end{equation}
In Eq.~(\ref{Gamma}), $\delta(\varepsilon,\varepsilon^\prime)$ denotes the Kronecker delta function, 
which accounts for the conservation of energy during collisions between fermions and bosons, 
and $n_\mbf{n}^b$ denote the occupation numbers of the bosonic reservoir. 
The operator $H_{F-F}$ describes an effective interaction between fermions, which is mediated by bosonic atoms:
\begin{equation}
H_{F-F} = \sum_{\mbf{n},\mbf{n}^{\prime},\mbf{m},\mbf{m}^{\prime}} 
\Delta_{\mbf{n},\mbf{n}^{\prime}}^{\mbf{m},\mbf{m}^{\prime}} 
a_\mbf{n}^{\dagger} a_{\mbf{n}^\prime} a_\mbf{m}^{\dagger} a_{\mbf{m}^\prime},
\end{equation}
where
\begin{equation}
\Delta_{\mbf{n},\mbf{n}^{\prime}}^{\mbf{m},\mbf{m}^{\prime}} = 
\delta(\varepsilon_\mbf{n}+\varepsilon_\mbf{m},\varepsilon_{\mbf{n}^\prime}+\varepsilon_{\mbf{m}^\prime})
\sum_{\mbf{k},\mbf{q}} U_{\mbf{n},\mbf{n}^{\prime},\mbf{q},\mbf{k}} 
U_{\mbf{m},\mbf{m}^{\prime},\mbf{k},\mbf{q}} n_\mbf{q}^b (1 + n_\mbf{k}^b)
\frac{1 - \delta(\varepsilon_\mbf{m}+\varepsilon_\mbf{k}^b,\varepsilon_{\mbf{m}^\prime}+\varepsilon_\mbf{q}^b)}
{\varepsilon_{\mbf{m}^\prime}+\varepsilon_\mbf{q}^b-\varepsilon_\mbf{m}-\varepsilon_\mbf{k}^b}.
\end{equation}
We assume that interactions between bosons and fermions are sufficiently small, 
and hence $H_{F-F}$ does not influence significantly the single-particle eigenstates in a harmonic trap.

To calculate the dynamics of a fermionic gas from the ME we introduce some further approximations. First, by
considering the decoherence induced by the interactions between bosons and fermions one can argue that the 
nondiagonal elements of the density matrix in the energy representation decay on a time-scale 
much shorter than the time scale of the dynamics due to sympathetic cooling \cite{Papenbrock}. 
Inclusion of $H_{F-F}$ lifts the degeneracy of the states of equal energy but different 
occupation numbers of single-particle eigenstates, and hence the decoherence affects also the nondiagonal terms with 
respect to the occupation numbers. Then, we apply ergodic approximation assuming that for a given energy all
diagonal elements of $\rho$ are equal. This requires that the process of equilibration within a single energy shell 
is much faster than for the whole system. Thus, the density matrix $\rho$ can be written in the form \cite{Papenbrock}
\begin{equation}
\label{rho}
\rho (t) = \sum_{E,\lambda} \frac{p_E(t)}{\Gamma(N_f,E)} \left|E,\{n_\mbf{n}\}_\lambda\right\rangle 
\left\langle E,\{n_\mbf{n}\}_\lambda \right|.
\end{equation}
Here, $\lambda$ enumerates microstates $\{n_\mbf{n}\}_\lambda$ of the fermionic component, and 
$\left|E,\{n_\mbf{n}\}_\lambda\right\rangle$ represents the state with energy $E$ and 
a given distribution of fermions. The probability that 
the system of fermions has energy $E$ is denoted by $p_E(t)$, while $\Gamma(N_f,E)$ is the number 
of microstates with energy $E$ and number of particles $N_f$. Substituting $\rho(t)$ given by Eq.~(\ref{rho}) 
into the ME~(\ref{ME}), and neglecting the terms due to the $H_{F-F}$, 
we obtain the equations determining the dynamics of the probabilities $p_E(t)$:
\begin{equation}
\label{pE}
\frac{d p_E}{d t} = 2 \sum_{\mbf{n},\mbf{l}} \Gamma_{\mbf{n},\mbf{m}}^{\mbf{m},\mbf{n}} \left(
p_{E+ \varepsilon_\mbf{n} - \varepsilon_\mbf{m}}(t) \left\langle n_\mbf{n} (1 - n_\mbf{m}) 
\right\rangle_{E+ \varepsilon_\mbf{n} - \varepsilon_\mbf{m}} - p_{E}(t) \left\langle n_\mbf{n} (1 - n_\mbf{m}) 
\right\rangle_{E} \right),
\end{equation}
where $ \left\langle n_\mbf{n} (1 - n_\mbf{m}) \right\rangle_E$ stands for the microcanonical average over 
microstates with energy $E$, and $n_\mbf{n}=a_\mbf{n}^{\dagger} a_\mbf{n}$. Probabilities $p_{E}(t)$
determine all the thermodynamic quantities of the gas of fermions. In particular, its mean energy is 
given by $\left\langle E \right\rangle = \sum_{E} E p_E(t)$. Combining this expression with Eq.~(\ref{pE}), we 
calculate the rate $\Gamma \equiv - \frac{d}{d t} \left\langle E \right\rangle$ of cooling of the fermions by 
the bosonic reservoir:
\begin{equation}
\label{dE}
\Gamma = 2 \sum_{\mbf{n},\mbf{m}} \Gamma_{\mbf{n},\mbf{m}}^{\mbf{m},\mbf{n}} \left(
\varepsilon_\mbf{n} - \varepsilon_\mbf{m} \right) \sum_{E} p_E(t) 
\left\langle n_\mbf{n} (1 - n_\mbf{m}) \right\rangle_{E}.
\end{equation}
Since we are mainly interested in the regime where fermions are sufficiently
cold and the effects of heating due to the losses become important, we assume that the gas of fermions 
is close to equilibrium and evaluate $p_E(t)$ from the canonical or 
the grand-canonical ensemble. We chose the latter possibility and postulate the grand-canonical 
distribution for both fermionic component and bosonic reservoir. Substituting Eq.~(\ref{Gamma}) for the rates $\Gamma_{\mbf{n},\mbf{m}}^{\mbf{m},\mbf{n}}$ into Eq.~(\ref{dE}), we arrive at
\begin{equation}
\label{GammaE}
\Gamma = \frac{2\pi}{\hbar^2 \omega_b} \sum_{\mbf{n},\mbf{m},\mbf{k},\mbf{q}}
\left|U_{\mbf{m},\mbf{n},\mbf{k},\mbf{q}}\right|^2 
n_\mbf{n} (1 - n_\mbf{m}) n_\mbf{q}^b (1 + n_\mbf{k}^b)
\left(\varepsilon_\mbf{n} - \varepsilon_\mbf{m} \right)
\delta(\varepsilon_\mbf{n}+\varepsilon_\mbf{q},\varepsilon_{\mbf{m}}+\varepsilon_{\mbf{k}}),
\end{equation} 
where $n_\mbf{n} = z e^{- \beta \varepsilon_\mbf{n}}/(1+z e^{- \beta \varepsilon_\mbf{n}})$,
$n_\mbf{q}^b = z_b e^{- \beta_b \varepsilon_\mbf{q}^b}/(1 - z_b e^{- \beta_b \varepsilon_\mbf{q}^b})$,
$\beta=1/(k_B T)$, $\beta_b = 1/(k_B T_b)$, and
$z$, $z_b$ denote the fugacities of fermions and bosons respectively. First we perform summation over 
degeneracies of the single-particle states. To this end we utilize the following approximate result 
\begin{equation}
\label{SumDeg}
\sum_{\mbf{n},\mbf{m},\mbf{k},\mbf{q}} \left|U_{\mbf{m},\mbf{n},\mbf{k},\mbf{q}}\right|^2 
\delta(\varepsilon_\mbf{n},e_1) \delta(\varepsilon_\mbf{m},e_2) 
\delta(\varepsilon_\mbf{q},e_3) \delta(\varepsilon_\mbf{k},e_4) \approx 
\frac{D(\mrm{min}(e_3,e_4)) g^2}{4 \pi^4 a_b^2 a_{\!f}^4},  
\end{equation}
where $a_b = \sqrt{\hbar/m_b \omega_b}$, $a_{\!f} = \sqrt{\hbar/m_{\!f} \omega_f{\!}}$, $D(e)$ denote the 
degeneracy of the energy shell $e$, and $\mrm{min}(e_3,e_4)$ represents the minimum of $e_3$ and $e_4$.
This formula can be derived by taking the semiclassical limit of Eq.~(\ref{GammaE}) and comparing 
with the semiclassical quantum Boltzmann equation in ergodic approximation \cite{Geist1,Luiten,Holland}. 
The summation formula is valid for 
$\mrm{min}(e_3,e_4)/(\hbar \omega_b) \leq \lambda^2 \mrm{min}(e_1,e_2)/(\hbar \omega_{\!f})$ and
$\mrm{min}(e_3,e_4)/(\hbar \omega_b) \leq \lambda^{-2} \mrm{min}(e_1,e_2)/(\hbar \omega_{\!f})$
where $\lambda = a_f/a_b$. For values of $\lambda$ which are not much larger or much smaller than one, 
this condition is fulfilled for collisions at sufficiently low temperatures, 
which take place close to the Fermi surface for fermions and  
close to the ground state for bosons. We have confirmed numerically the validity of formula (\ref{SumDeg}),
and verified that it is quite accurate even for the lowest
states of the harmonic potential. Substituting (\ref{SumDeg}) into Eq.~(\ref{GammaE}) yields
\begin{equation}
\label{GammaE1}
\Gamma = \frac{2 A \omega_b }{\pi} \sum_{e_1,e_2,e_3,e_4} 
n_{e_1} (1 - n_{e_2}) n_{e_3}^b (1 + n_{e_4}^b) (e_1 -e_2) D(\mrm{min}(e_3,e_4))
\delta(e_1+e_3,e_2+e_4),
\end{equation} 
where $n_{e_1} = z e^{- \beta e_1}/(1+z e^{- \beta e_1})$ and 
$n_{e_3}^b = z_b e^{- \beta_b e_3}/(1 - z_b e^{- \beta_b e_3})$, and 
$A = m_b^2 a^2 a_b^2/(\mu_m^2 a_{\!f}^4)$.
For a sufficiently large system, the summation over $e_1,e_2,e_3,e_4$ can be replaced by an integration, 
which corresponds to taking the thermodynamic limit. This procedure is correct with respect to the collisions 
with the thermal part of the bosonic component. In the case of a Bose-condensed reservoir, 
the single-particle ground state requires a separate
treatment. Let us first consider the rate $\Gamma_\mrm{th}$ due to the interactions with the thermal atoms in the 
reservoir. Replacing sums by integrals, after some algebra we obtain 
\begin{equation}
\label{Gth}
\Gamma_\mrm{th} = \frac{4 A}{\pi}
\frac{ (k_B T)^3 (k_B T_b)^3}{\hbar^5 \omega_{\!f}^2 \omega_b^2}
\sum_{l=0}^{\infty} \zeta(3,1+l) \left[ \zeta\left(3,{\tst \frac{T}{T_b} l + 1}\right) 
- \zeta\left(3,{\tst \frac{T}{T_b}(1+l)}\right) \right],
\end{equation}
which is valid for temperatures of fermions much smaller than the Fermi temperature $T_F$, 
and under assumption that the reservoir is 
Bose-condensed. Here $\zeta(s,a)$ denotes the Hurwitz zeta function \cite{Elizalde}.
We note that $\Gamma_\mrm{th}$ does not depend on the number of fermions. This is  
is related with the fact that at low temperatures the collisions occur only close to the Fermi surface.
On the contrary, taking into account that the number of the thermal atoms in a trapped Bose-condensed gas is 
$N_\mrm{th}=\zeta(3)(k_B T_b/(\hbar \omega_b))^3$, we observe that $\Gamma_\mrm{th}$ is proportional to $N_\mrm{th}$.
When the temperature of both components is equal, we expect that there will be no energy exchange between bosons and
fermions and, as can be easily verified, this feature is properly taken into account by Eq.~(\ref{Gth}).
To calculate the rate $\Gamma_{0}$ describing the cooling due to collisions with the condensed  
part of the reservoir, we replace sums by integrals in Eq.~(\ref{GammaE}), putting $e_3=0$ or $e_4 =0$. This yields
\begin{equation}
\label{G0}
\Gamma_0 = \frac{4 \zeta(3) A}{\pi} \frac{N_0 \omega_b}{(\hbar \omega_{\!f})^2}
\left[ (k_B T)^3 - (k_B T_b)^3 \right]. 
\end{equation}
We note that at low temperatures the energy exchange with the condensate is proportional to the number of 
condensed atoms $N_0$, but does not depend on the number of fermions.

We turn now to the description of the heating due to particle losses in the degenerate Fermi gas.
The losses create holes in the Fermi sea, which after subsequent thermalization 
increase the temperature of the system \cite{Timmermans}.
We assume that particles are removed with a rate $\gamma$, which  
is the same for all single-particles states: $\frac{d}{dt}\left\langle n_\mbf{n} \right\rangle = -\gamma
\left\langle n_\mbf{n} \right\rangle$,
where $\left\langle n_\mbf{n} \right\rangle$ denotes the mean occupation number of the state $\mbf{n}$. 
For background losses, the rate $\gamma$ does not depend on the density of fermions $n_f$, while in the case of two- and
three-body losses it is proportional to $n_f$ and $n_f^2$, respectively. The removal process
does not change the mean energy per particle: $(\langle E \rangle /N_f) (t) =$~const. At temperatures $T/T_F \ll 1$,  
$\langle E \rangle/N_f = 3 E_F(N_f)/4 + (\pi^2/2) k_B^2 T^2 /E_F(N_f)$ with the
Fermi energy $E_F(N_f) \approx \hbar \omega_{\!f} (6 N_f)^{1/3}$. The creation of holes, 
can be equivalently described as a process that heats the system, but, in first approximation, does not 
change the number of particles. We obtain that the rate of heating is given by 
\begin{equation}
\label{GH}
\Gamma_\mrm{H} = N_f \gamma E_F \left[ {\tst \frac 14 - \frac{\pi^2}{6}\left(T/T_F\right)^2 }\right],
\end{equation}
where we neglect terms of order higher than $(T/T_F)^2$.

In the absence of particle losses, the Fermi component is cooled down until it reaches the temperature of 
the bosonic reservoir $T_b$. The heating produced by the hole creation, however, prevents from reaching $T_b$,
and sets a constraint on the lowest temperature achievable with the sympathetic colling.
The final temperature of the Fermi gas $T_\mrm{fin}$, at which 
the cooling is balanced by the hole heating, is determined by the condition:
$\Gamma_\mrm{th}+\Gamma_0=\Gamma_\mrm{H}$. When the losses are sufficiently small, 
the final temperature of fermions $T_\mrm{fin}$ only slightly differs 
from $T_b$, and in this case its value can be estimated analytically. To this end, we perform a Taylor expansion of 
the cooling rate $\Gamma = \Gamma_\mrm{th}+\Gamma_0$ at $T = T_B$, keeping only the lowest order term 
$\Gamma (T) \approx (\partial \Gamma/(\partial T))|_{T=T_b}(T-T_b)$. This yields
\begin{equation}
\label{GA1}
\Gamma \approx \frac{12 A}{\pi}
\frac{\omega_b(k_B T_b)^3}{(\hbar \omega_{\!f})^2}
\frac{T-T_b}{T_b} \left[\zeta(3) N_0 + c \left(\frac{k_B T_b}{\hbar \omega_b}\right)^3 \right],
\end{equation}
which is valid for $|T-T_b| \ll T_b$. Here $c$ is a numerical constant given by 
$c = \sum_{l=0}^{\infty} \zeta(3,1+l) \zeta(4,1+l) \simeq 1.3196$.
For temperatures of the bosonic reservoir $T_b$ much smaller than the critical temperature $T_C$, one can
neglect $c (k_B T_b/(\hbar \omega_b))^3$ in comparison to $\zeta(3) N_0$, and in this regime Eq.~(\ref{GA1}) reduces to 
\begin{equation}
\label{GA2}
\Gamma \approx 
\frac{12 A}{\pi}
\frac{\hbar \omega_b^4}{\omega_{\!f}^2} \left(\frac{T_b}{T_C}\right)^3
\frac{T-T_b}{T_b} {N_0}^2.
\end{equation}
This formula combined with the rate $\Gamma_\mrm{H}$, which at $T \ll T_F$ can be approximated by 
$\Gamma_\mrm{H} \approx N_f \gamma E_F/4$, 
leads to the following result for the final temperature of the Fermi gas
\begin{equation}
\label{Tfin}
T_\mrm{fin} = T_b \left(1 + \frac{6^{1/3} \pi}{48 A} \frac{\gamma \omega_{\!f}^3}{\omega_b^4}
\frac{N_f^{4/3}}{N_0^2} \left(\frac{T_C}{T_b}\right)^3 \right).
\end{equation}

The lowest temperature in the sympathetic cooling of a Fermi gas is achieved when the reservoir is totally condensed.
In this particular case, the final temperature can be also calculated analytically. We observe that at $T_b = 0$,
$\Gamma_\mrm{th}=0$, while for $T \ll T_F$  the rate of loss-induced heating 
can be approximated by $\Gamma_\mrm{H} \approx N_f \gamma E_F/4$. This yields the final temperature of fermions:
\begin{equation}
\label{Tfin0}
T_\mrm{fin} = b T_F \left( \frac{\gamma N_f^{1/3}}{A \omega_b N_0} \right)^{1/3},
\end{equation}
where $b = ( 6^{1/3} \pi/(96 \zeta(3)))^{1/3} \simeq 0.367$. We note that $T_\mrm{fin}$ very weakly depends on
the number of fermions $N_f$, and is mainly determined by the efficiency of the cooling process (through 
parameters $N_0$, $A$, $\omega_b$), and by the loss rate $\gamma$.
%%%%%%%%%%%%%%%%% Figure 01 %%%%%%%%%%%%%%%%%%%%%%
\begin{figure}
	 \twofigures[height=5.4cm]{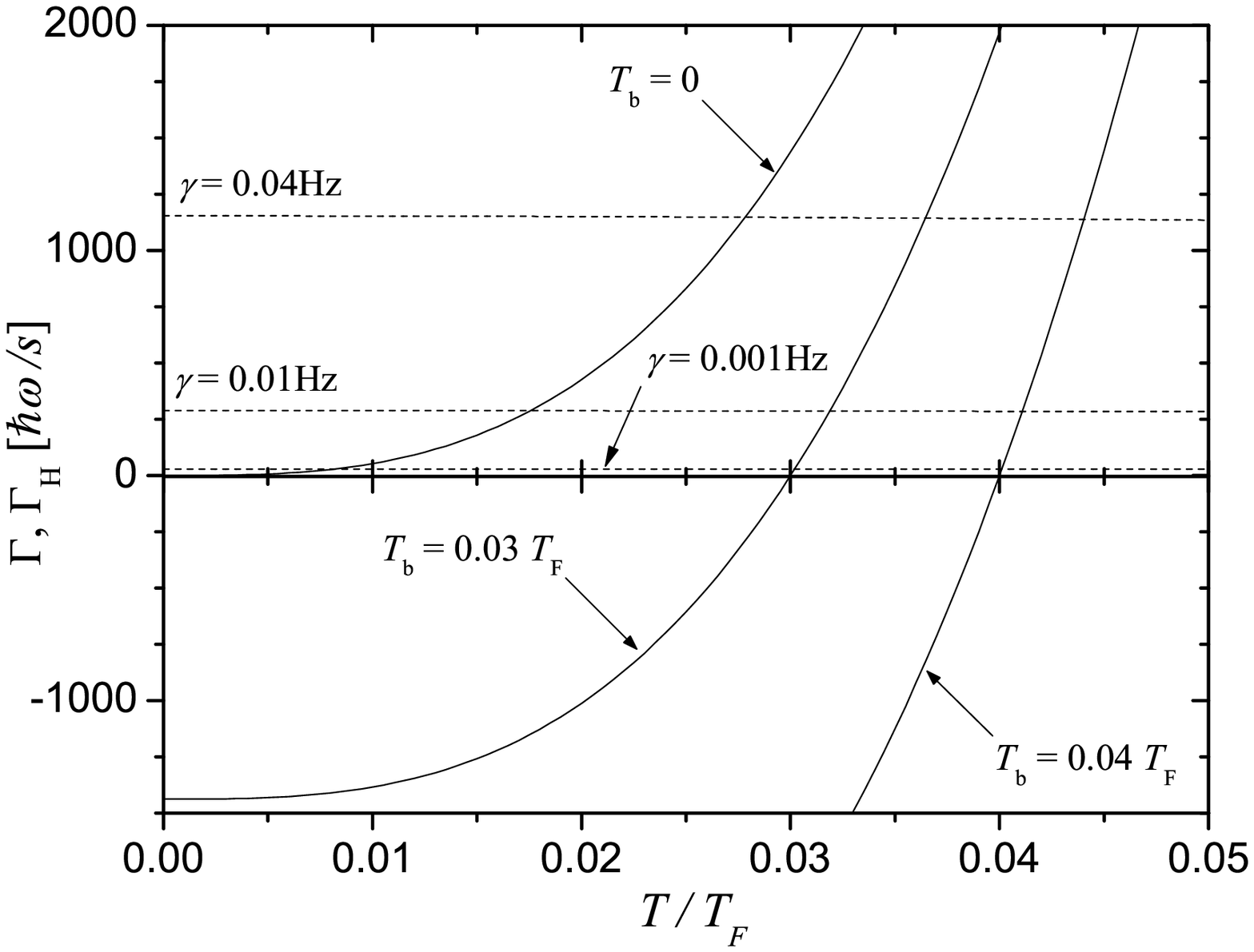}{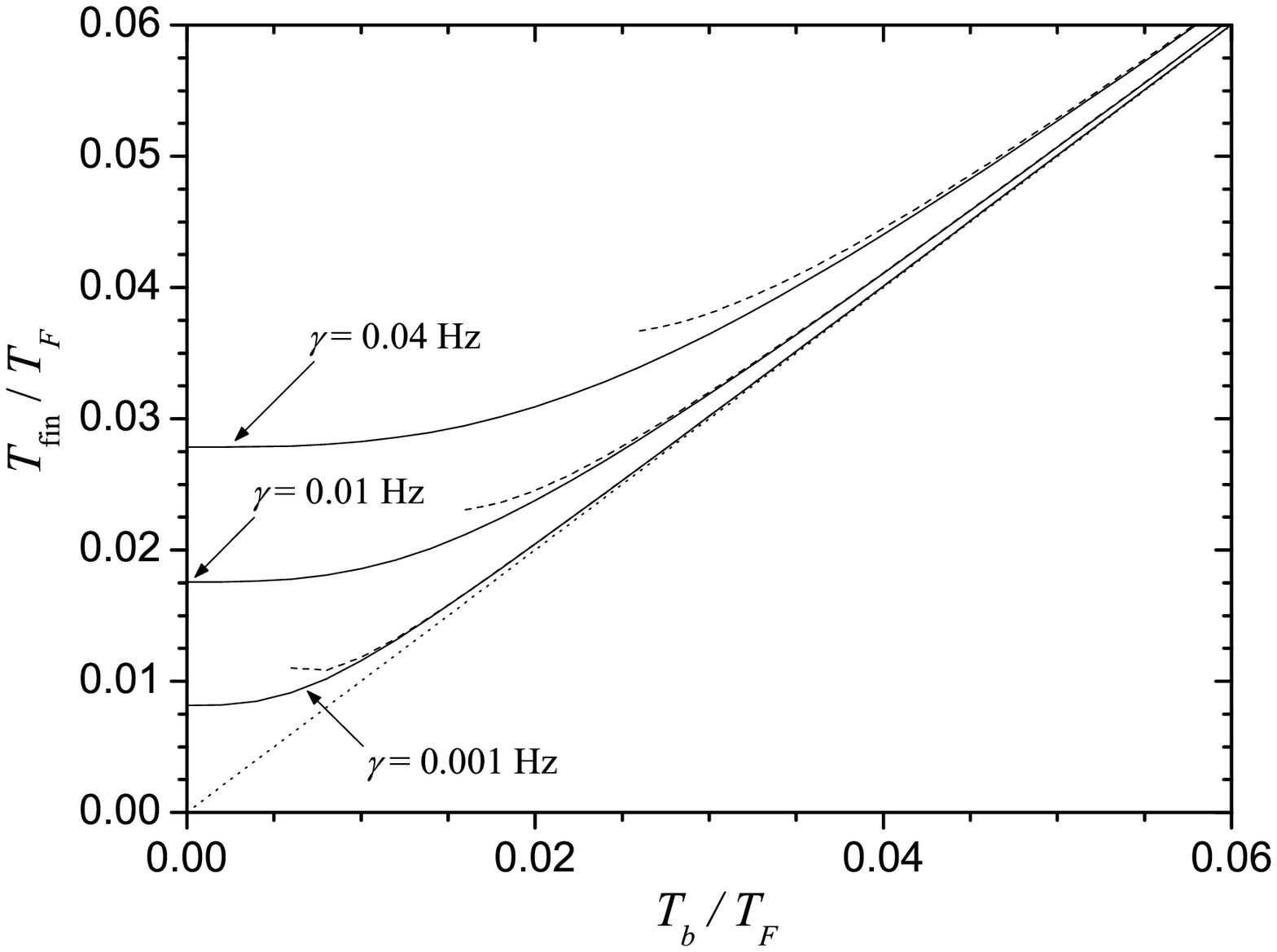}
	 \caption{
	 \label{fig:Gammy}
	 Cooling rates $\Gamma$ (solid lines) and heating rates $\Gamma_{\mrm{H}}$ 
	 (dashed lines) versus the fermion 
temperature $T$. The rates were calculated for three different temperatures 
of the
	 bosonic reservoir: $T_b=0$, $0.03 T_F$, $0.04 T_F$, and for three different loss rates 
$\gamma = 0.04$Hz, $0.01$Hz, and $0.001$Hz. The values of the other parameters correspond 
	 to the experiment described in \cite{Salomon} (see text for details). }
	 \caption{
	 \label{fig:Tfin}
	 Final temperature of the Fermi gas versus the temperature of the bosonic reservoir $T_b$ calculated for
	 the parameters of Fig.~\ref{fig:Gammy}.
	 The plot shows the numerical values evaluated from the rates 
	 given by Eqs.~(\ref{Gth}), (\ref{G0}), and (\ref{GH}) (solid lines), the approximate result of
	 Eq.~(\ref{Tfin}) (dashed lines), and the line $T=T_b$, which is the final temperature in 
	 the absence of losses (dotted line). }
\end{figure}
%%%%%%%%%%%%%%%%%%%%%%%%%%%%%%%%%%%%%%%%%%%%%%%%%%%

To estimate the influence of the loss-induced heating on the final temperature of fermions,
we have performed numerical calculations for the mixture of bosonic ${}^{7}$Li with fermionic ${}^{6}$Li 
\cite{Salomon,Hulet}. 
In our calculations we assume: $N_b =1.5 \times 10^4$, $N_{\! f} =4 \times 10^3$, 
$\omega_b = \omega_{\! f} = 2 \pi \times 1270$Hz, and $a = 2.0$nm, which corresponds to the experimental parameters of
Ref.~\cite{Salomon}. Fig.~\ref{fig:Gammy} shows, as a function of the temperature 
of the fermions, the cooling rates $\Gamma = \Gamma_0 + \Gamma_\mrm{th} $ evaluated from 
Eqs.~(\ref{Gth}) and (\ref{G0}) for three different temperatures of the bosons:
$T_b=0$, $0.03 T_F$, and $0.04 T_F$ (solid lines). The dashed lines show the heating rate $\Gamma_\mrm{H}$ 
given by Eq.~(\ref{GH}). Since the loss rate $\gamma$ depends on the particular experimental conditions,  
we have used three different values of $\gamma$: $0.04$Hz, $0.01$Hz and $0.001$Hz.
The final temperature of fermions is determined by the crossing of $\Gamma$ and $\Gamma_\mrm{H}$.
We observe that for the totally condensed reservoir ($T_b = 0$) and for fermion temperatures 
$T \lesssim 0.01 T_C$, the cooling rates become very small due to Pauli blocking. In this regime 
the presence of even very small losses prevents from reaching lower temperatures. For higher temperatures
of the bosonic component ($T_b = 0.03 T_F$ and $T_b = 0.04 T_F$), the hole heating shifts only slightly 
the final fermion temperature with respect to $T_b$, and $T_\mrm{fin}$ 
can be determined by expanding the cooling rate around $T_b$ as in the case of Eq.~(\ref{Tfin}).
The dependence of the final fermion temperature $T_\mrm{fin}$ on $T_b$ is shown in Fig.~\ref{fig:Tfin}, where  
the results calculated from the condition $\Gamma_\mrm{th}+\Gamma_{0}=\Gamma_\mrm{H}$, using Eqs.~(\ref{Gth}), 
(\ref{G0}) and (\ref{GH}), are compared with Eq.~(\ref{Tfin}). 

In our analysis we have neglected the interactions within the bosonic reservoir. This can be a good
approximation for the conditions of experiment \cite{Salomon}, where the condensate is relatively small 
($N_b \sim 10^4$) and the scattering length for ${}^{7}$Li atoms in the state $|F=1,m_F = -1\rangle$
is $a=0.27$nm. In general, however, the interactions modify the density profile of the bosonic cloud,
and introduce the quantum depletion of the condensate. For repulsive forces
the size of an interacting condensate is much larger in comparison to the ideal-gas. This can 
reduce the collisional rate between bosons and fermions and consequently decrease the cooling efficiency.
On the other hand, the quantum depletion reduces the number of condensed atoms,
transfering them into excited states, which increases the effective temperature of the reservoir.
Therefore, the most effective cooling is achieved for noninteracting bosons, 
which can be eventually realized with the help of Feshbach resonances.

Summarizing, we have investigated the sympathetic cooling of a 
trapped Fermi gas in contact with a Bose-condensed
reservoir, and  shown that the loss-induced heating limits the 
cooling at low temperatures. In the 
regime of weak heating and in the case of totally condensed Bose gas,
we have derived the analytical results for the final temperature of the fermions. 

\acknowledgements
We acknowledge fruitful discussions with L. Carr. Z.I. acknowledges BEC-INFM for hospitality and financial support. 
This work has been supported by 
the Deutsche Forschungsgemeinschaft (SFB 407, SPP 1116, 432 POL), ESF Program BEC2000+,
and by the Alexander von Humboldt foundation.

\end{document}